\documentclass[oneside,12pt]{article}

\usepackage{color}
\usepackage{greekbf}
\usepackage{float}
\usepackage{amsmath,amssymb}
\usepackage{enumerate}
\usepackage{subfigure}
\usepackage{booktabs}

\usepackage{mathtools}
\usepackage{graphicx}
\usepackage{hyperref}
\hypersetup{bookmarks=false,
colorlinks=true,
linkcolor=black,
urlcolor=black,
pagebackref=true,
pdfstartview={FitH},
bookmarksopenlevel=2
}

\newcommand{\tB}[1]{{\color{blue} #1}}
\newcommand{\tG}[1]{{\color{green} #1}}
\newcommand{\tO}[1]{{\color{orange} #1}}

\definecolor{light-gray}{gray}{0.95}

\newtheorem{remark}{Remark}

\definecolor{red}{rgb}{0.9,0,0}
\definecolor{blue}{rgb}{0.235,0.114,0.976}
\definecolor{green}{rgb}{0.43,0.227,0.357}
\definecolor{orange}{rgb}{0.96,0.45,0.184}

\DeclareMathAlphabet{\mathbf}{OT1}{cmr}{bx}{it}

\definecolor{red}{rgb}{0.9,0,0}
\definecolor{blue}{rgb}{0.2,0.2,0.8}
\definecolor{green}{rgb}{0.0,0.5,0.2}
\definecolor{darkblue}{rgb}{0.2,0.2,0.5}
\definecolor{orange}{rgb}{1,0.5,0}

\newlength{\lengthD}
\newcommand{\bydef}{\,\raise.050ex\hbox{\rm:}\kern-.025em\hbox{\rm=}\,}
\newcommand{\defby}{=\raise.075ex\hbox{\kern-.325em\hbox{\rm:}}\,}

\def\qed{\relax\ifmmode\hskip2em \Box\else\unskip\nobreak\hskip1em $\Box$\fi}
\newcommand {\Dc}  {\mathcal{D}}
\newcommand {\Ec}  {\mathcal{E}}

\newcommand {\Uc}  {\mathcal{U}}

\newcommand {\db} {\mathbf{d}}
\newcommand {\eb} {\mathbf{e}}

\newcommand {\pb} {\mathbf{p}}

\newcommand {\xb} {\mathbf{x}}

\newcommand {\ub} {\mathbf{u}}

\newcommand {\Ab} {\mathbf{A}}

\newcommand {\Eb} {\mathbf{E}}

\newcommand {\Kb} {\mathbf{K}}

\newcommand {\Ro} {\mathbb{R}}

\newcommand {\tr}[1]{\mbox{tr}\, #1}

\usepackage{epstopdf}
\usepackage{import}
\graphicspath{{Figures/}}

\DeclareMathOperator{\Thc}{\Theta}
\DeclareMathOperator{\Ths}{\overset{(s)}{\vartheta}}

\begin{document}

\title{\vspace{-1cm} \textbf{A 2D metamaterial \\with auxetic out-of-plane behavior\\ and non-auxetic in-plane behavior}}

\author{
Cesare Davini$^1$ \!\!\!\!\! \and Antonino Favata$^2$ \!\!\!\!\! \and Andrea Micheletti$^3$\!\!\!\!\!  \and Roberto Paroni$^4$
}

\date{\today}

\maketitle

\vspace{-1cm}
\begin{center}
{\small
$^1$ Via Parenzo 17, 33100 Udine\\
\href{mailto:cesare.davini@uniud.it}{cesare.davini@uniud.it}\\[8pt]
$^2$ Department of Structural and Geotechnical Engineering\\
Sapienza University of Rome, Rome, Italy\\
\href{mailto:antonino.favata@uniroma1.it}{antonino.favata@uniroma1.it}\\[8pt]

$^3$ Dipartimento di Ingegneria Civile e Ingegneria Informatica\\
University of Rome Tor Vergata, Rome, Italy\\
\href{mailto:micheletti@ing.uniroma2.it}{micheletti@ing.uniroma2.it}\\[8pt]

$^4$ Dipartimento di Architettura, Design e Urbanistica\\
University of Sassari, Alghero (SS), Italy\\
\href{mailto:paroni@uniss.it}{paroni@uniss.it}
}
\end{center}

\pagestyle{myheadings}

\vspace{-0.5cm}
\section*{Abstract}
Customarily, in-plane auxeticity and synclastic bending behavior (i.e. out-of-plane auxeticity) are not independent, being the latter a manifestation of the former. Basically, this is a feature of three-dimensional bodies. 
At variance, two-dimensional bodies have more freedom to deform than three-dimensional ones. Here, we exploit this peculiarity and propose a two-dimensional honeycomb structure with out-of-plane auxetic behavior opposite to the in-plane one. With a suitable choice of the lattice constitutive parameters, in its continuum description such a structure can achieve the whole range of values for the bending Poisson coefficient, while retaining a membranal Poisson coefficient equal to 1. In particular, this structure can reach the extreme values, $-1$ and $+1$, of the bending Poisson coefficient. Analytical calculations are supported by numerical simulations, showing the accuracy of the continuum formulas in predicting the response of the discrete structure.

\vspace{1cm}

\noindent {\bf Keywords}: Auxetic materials, metamaterials, bending Poisson coefficient, synclastic surfaces.

\tableofcontents

\section{Introduction}

When materials with positive Poisson ratio are uniaxially stretched, they shrink in directions transversal to the applied load and when compressed they expand transversally. For this reason, in a plate regarded as a thin three-dimensional body the bending in one direction induces a bending with an opposite curvature in the transversal direction.
	This {\it anticlastic behavior} of a plate is well known and is simply explained by observing that the ``main'' bending produces compressions in, say, the upper half of the plate and tractions in the lower part. Thence, the upper part tends to expand transversally, while the lower part tends to shrink, and this is naturally achieved with a bending in the transversal direction with a curvature of opposite sign with respect to that in the ``main'' bending direction.

Materials with negative Poisson ratio are called {\it auxetic} and have a behavior that is ``opposite'' to that of {\it non-auxetic} materials, described above. In particular, auxetic materials have a {\it synclastic plate behavior}. 

For three-dimensional bodies, the sign of the Poisson ratio determines the synclastic or anticlastic plate behavior. This is not the case for  two-dimensional bodies deforming in the three-dimensional space: roughly speaking, because there is no upper and  lower region that is compressed or stretched due to bending.  In two-dimensional bodies the in-plane deformations yield situations that are naturally related to those of a classical continuum, and one may define a Poisson ratio as the negative of the ratio of the transversal strain to the axial strain for a uniaxial stress state. Such a ratio was called {\it membranal Poisson coefficient} in \cite{Davini_2017a}. 
	Thus, as explained above, the membranal Poisson coefficient does not determine the synclastic/anticlastic plate behavior. 
	For the out-of-plane deformations of a two-dimensional body, on the contrary, there is no natural resemblance with what occurs in mechanics of ordinary continua; for instance, it is necessary to account for the dependence of the mechanical response upon higher order strain measures,  {\it e.g.}, upon curvatures. Therefore, the introduction of a Poisson coefficient that applies to this class of deformations needs an independent definition. Here, we choose to define it as minus the ratio between the principal curvatures in states of uniaxial bending and call it {\it bending Poisson coefficient}, as done in \cite{Davini_2017a}. Hence a positive (negative) bending Poisson coefficient is synonymous of anticlastic (synclastic) plate behavior.
	
	By deducing the model from a micro-mechanic approach, see for instance \cite{Davini_2017} for the case of graphene, it can be shown that membranal and bending Poisson coefficients do describe different mechanical properties and can be regarded as distinct material parameters.   
	Then, with two types of Poisson coefficients it is natural to define two auxetic classes: two-dimensional materials with a negative membranal Poisson coefficient are said to have an {\it auxetic in-plane behavior}, and materials with a negative bending Poisson coefficient are said to have an {\it auxetic out-of-plane behavior.}
	
As the title pre-announces, we here show how to design a two-dimensional material with a microstructure that leads to a non-auxetic in-plane behavior and an auxetic out-of-plane behavior. Indeed, we  show that by appropriately tuning two material parameters, we may achieve all the possible values that the bending Poisson coefficient may take. To the best of our knowledge this is the first designed material with a non-auxetic in-plane behavior that has a synclastic plate behavior.

The first experimental evidence on the existence of 3D materials with a negative Poisson ratio was given for cubic crystals of pyrite by Love in 1927 \cite{Love_1927}. Other natural materials with a negative Poisson ratio include silicates \cite{Yeganeh_1992}, zeolites \cite{Grima_2000,Grima_2007}, and ceramics \cite{Song_2008}. The term auxetic was introduced in \cite{Evans_1991}, while
Lakes \cite{Lakes_1987} was the first to design an auxetic material consisting of a reentrant foam.
Since then,  materials with a negative Poisson ratio have been thoroughly analysed and designed: we refrain from referring to the huge literature and we limit ourselves to cite a few recent papers and reviews \cite{Evans_2000,Greaves_2011,Prawoto_2012,Zhang_2013,Cabras_2014,Lim_2015,Cabras_2016}.

Most often, auxetic materials are \textit{metamaterials}, namely engineered materials whose peculiar properties are determined by the geometric structure rather than the intimate constituent of the matter.
This kind of materials has received a great deal of attention for their importance in the development of the next generation of actuators, sensors, and smart responsive surfaces \cite{Sidorenko_2007,Bertoldi_2010}.

The microstructure that we propose is reminiscent of, and inspired by, that of graphene~--~a  two-dimensional material having a honeycomb pattern.
Generally speaking,  honeycomb structures have huge engineering applications thanks to their low density and high stiffness \cite{Bitzer_1997,Vinson_1999,Cutler_2005}. At variance with the common honeycomb structures, the lattice material that we consider is composed of pin-jointed rigid sticks arranged in a hexagonal pattern, pairwise connected by angular springs; each stick is the center of two C-shaped sequences of sticks  with  dihedral springs, storing energy whenever a change of the dihedral angles  spanned by the C-shaped sequence occurs. The presence of dihedral interactions is the first peculiar character of our material and descends from the study of the intimate composition of graphene; in this material, this interaction has indeed a crucial role in the development of its extraordinary mechanical properties \cite{Lu_2009,Favata_2016}.

The second peculiar character of the material we propose stems form the observation that graphene  does not have a configuration at ease. Indeed, the angular springs are supposed to be not stress-free in the regular hexagonal pattern: we take this into account by introducing a so-called \textit{wedge self-stress}.  Hence,  we have here three lattice material parameters: the elastic constant of the angular springs $k^\vartheta$, the elastic constant of the dihedral springs $k^\Theta$, and the wedge self-stress $\tau_0$. As shown in \cite{Davini_2017}, the constant
$k^\vartheta$ has influence only on the in-plane behavior, while  $k^\Theta$ and $\tau_0$ determine the bending stiffness and the bending Poisson coefficient. At variance with graphene, where the bending Poisson coefficient is determined by the chemical-physical interactions and turns out to be positive \cite{Davini_2017a}, in our metamaterial, by tuning the value of $k^\Theta$ and $\tau_0$ in the manufacturing process, the bending Poisson coefficient  is allowed to take all the admissible values, in particular the extreme ones $-1$ and $+1$.

The paper is organized as follows. In Section \ref{plates} we describe the energies of a plate-like 2D body, introduce the notion of membranal and bending Poisson coefficient and provide for them the constraints they have to satisfy. In Section \ref{microstructure} we introduce the microstructure of the metamaterial we propose, and we define their energetic membranal and bending contributions.  In Section \ref{Poissons} we determine the continuum limits of both  Poisson coefficients. In Section \ref{numerical} we collect some numerical results validating our theory, by adopting a computer code \textit{ad hoc} developed.

\section{Poisson coefficients in plate theory}\label{plates}

In this section we introduce the energies governing the in-plane and the out-of-plane displacements
of a body occupying a two-dimensional region  $\Omega$ and deforming in the three-dimensional space. Hereafter, this body is simply called {\it plate}. Besides introducing the notation used throughout the paper we  discuss the constraints that the material parameters have to satisfy.

\subsection{In-plane displacements for two-dimensional bodies}
The in-plane energy of a homogeneous and isotropic elastic plate, occupying the region $\Omega\subset\Ro^2$, is defined by 
\begin{equation}\label{Um}
\Uc^{(m)}(\ub)=\Ec\int_\Omega\nu^{(m)} (\mbox{tr}\, \Eb)^2+(1-\nu^{(m)})|\Eb|^2\,d\xb,
\end{equation}
where $\ub$ is  the in-plane infinitesimal displacement,
$\Eb=1/2 (\nabla\ub+\nabla\ub^T)$ is the strain, $\Ec$ is the \textit{stretching stiffness}, and
$\nu^{(m)}$ is the {\it membranal Poisson coefficient}.
One easily checks that the energy density $Q^{(m)}:\Ro^{2\times 2}_{\rm sym}\to \Ro$ given by
$$
Q^{(m)}(\Ab)=\Ec\Big(\nu^{(m)} (\mbox{tr}\, \Ab)^2+(1-\nu^{(m)})|\Ab|^2\Big)
$$
is strictly positive definite, {\it i.e.}, $Q^{(m)}(\Ab)>0$ for every
$\Ab\ne \mathbf{0}$,
if and only if
\begin{equation}\label{constraintm}
\Ec>0\qquad\mbox{and}\qquad -1<\nu^{(m)}<1.
\end{equation}

{\it The well-known membranal Poisson coefficient is the
	negative of the ratio of the  transversal strain to the axial strain for a uniaxial stress state}.

\subsection{Out-of-plane displacements for two-dimensional bodies}
The out-of-plane energy of the plate, supposed isotropic and linearly elastic, can be given the general form
\begin{equation}\label{Ub}
\Uc^{(b)}(w)=\frac 12 \Dc\int_\Omega (\Delta w)^2-2(1-\nu^{(b)})\det \nabla^2 w\,d\xb,
\end{equation}
where $w$ is the out-of-plane displacement, $\Dc$ is the bending stiffness, and $\nu^{(b)}$ is the {\it bending Poisson coefficient}.
Denoting by $Q^{(b)}:\Ro^{2\times 2}_{\rm sym}\to \Ro$ the quadratic form
$$
Q^{(b)}(\Kb)=\frac 12 \Dc\Big((\tr \Kb)^2-2(1-\nu^{(b)})\det \Kb\Big),
$$
we may also write
$$
\Uc^{(b)}(w)=\int_\Omega Q^{(b)}( \nabla^2 w)\,d\xb.
$$

The total energy of the plate has a minimizer, in the appropriate space and under the action of regular forces and suitable boundary conditions, if and only if the quadratic form $Q^{(b)}$ is strictly positive definite. One  verifies that this condition holds if and only if
\begin{equation}\label{constraint}
\Dc>0\qquad\mbox{and}\qquad -1<\nu^{(b)}<1.
\end{equation}

To understand the physical meaning of the bending Poisson coefficient, take $\Omega$ to be a rectangle and the $x$ and $y$ axes along the edges of the rectangle:
$$
\Omega=(0,a)\times (0,b),
$$
with $a$ and $b$ two positive constants denoting the side lengths of the rectangle. We denote by $M_x$ the bending moment per unit length acting on the edges parallel to the $y$ axis, {\it i.e.}, acting on $\{0\}\times (0,b)$ and on $\{a\}\times (0,b)$. Then, up to an infinitesimal rigid translation, the displacement that the plate undergoes is:
\begin{equation}\label{wMx}
w(x,y)=\frac{-M_x}{2\Dc(1-(\nu^{(b)})^2)}(x^2-\nu^{(b)} y^2),
\end{equation}
see equation $(d)$ of section 11 of \cite{Timoshenko_1959}.
From \eqref{wMx} we infer that
\begin{equation}\label{nub}
\nu^{(b)}=-\frac{\partial_{yy}w}{\partial_{xx}w},
\end{equation}
that is: {\it the bending Poisson coefficient is the
	negative of the ratio of the transversal curvature to the axial curvature for a uniaxial bending state}.

\begin{figure}[htbp]
	\centering%
	\subfigure[\label{fig:anticlastic}]%
	{\includegraphics[scale=0.5]{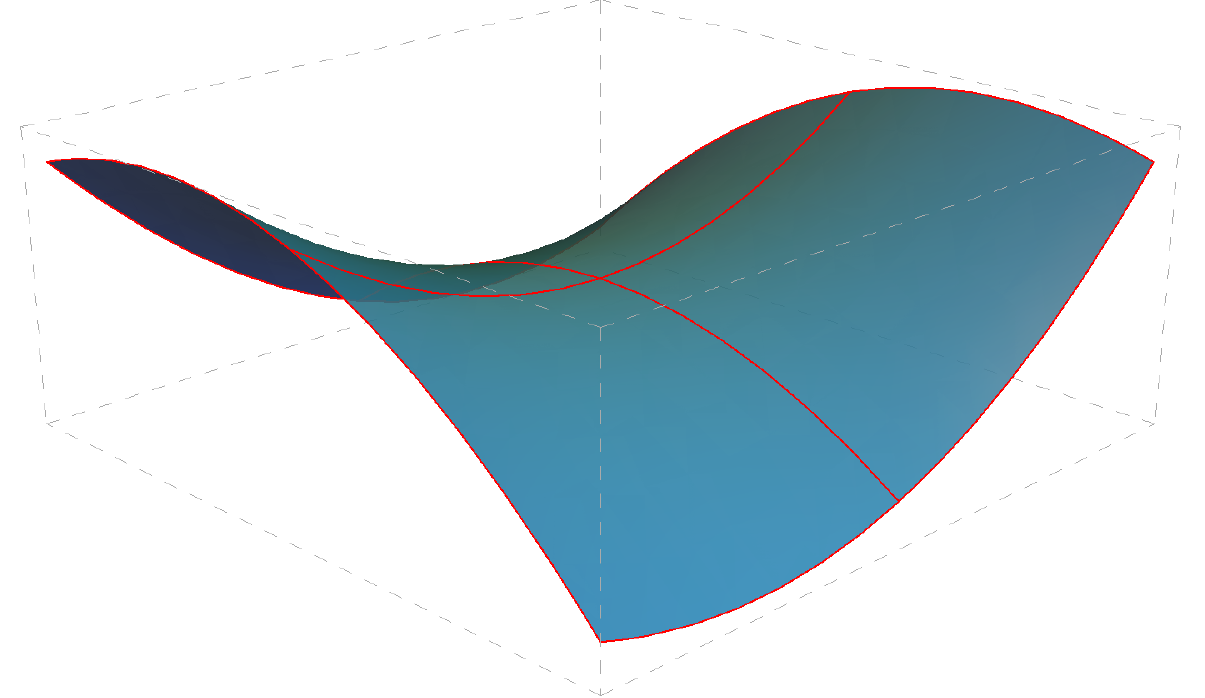}}\qquad\qquad
	\subfigure[\label{fig:monoclastic}]%
	{\includegraphics[scale=0.5]{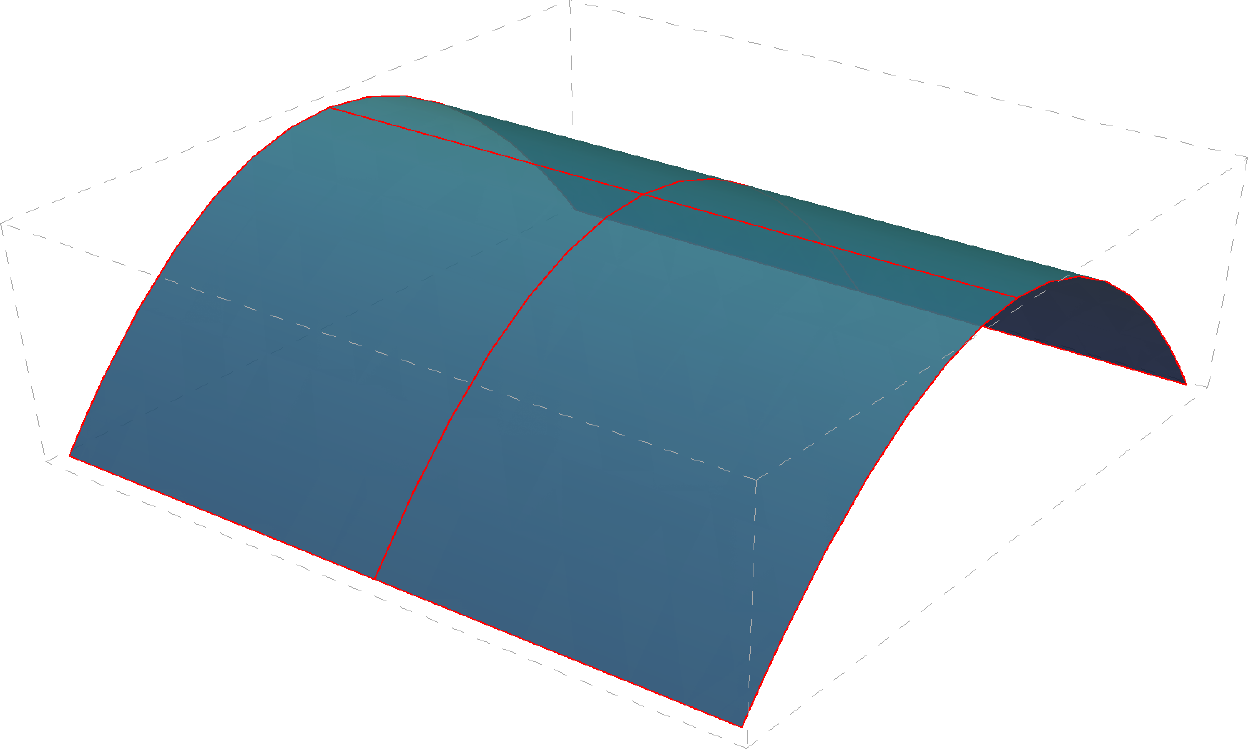}}\qquad\qquad
	\subfigure[\label{fig:synclastic}]%
	{\includegraphics[scale=0.5]{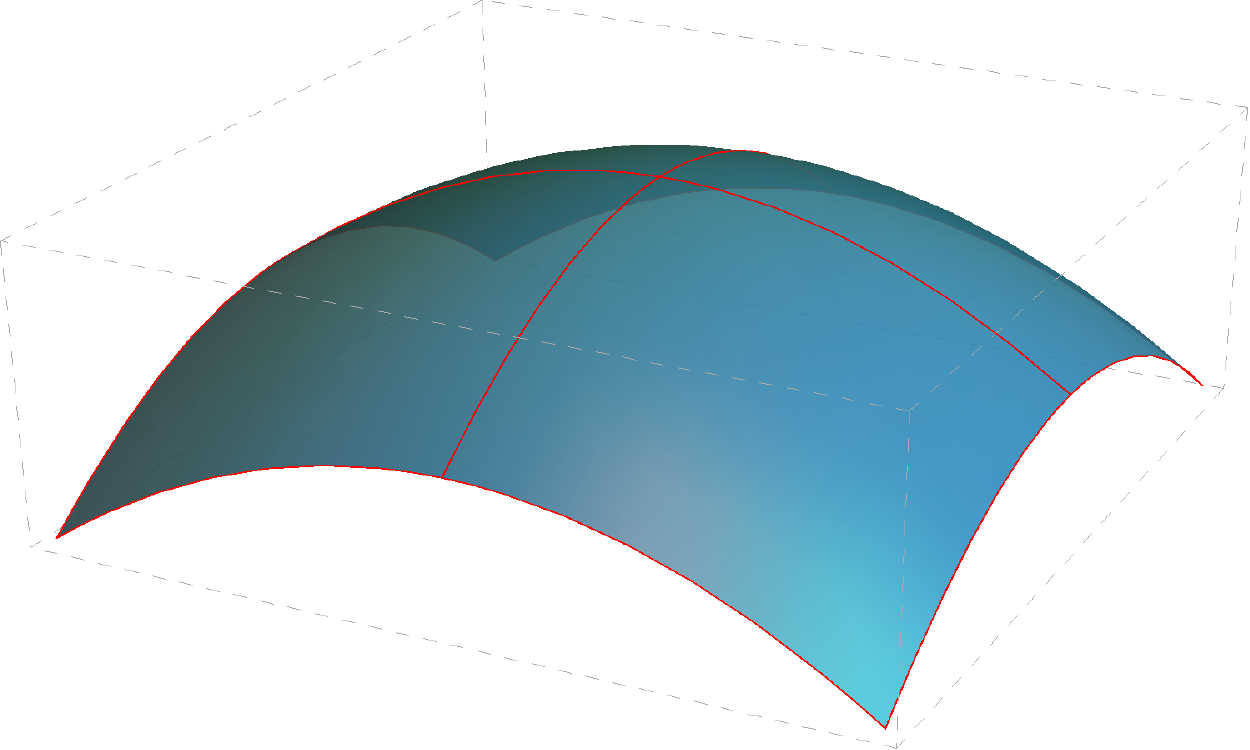}}
	\caption{(a) Anticlastic surface; (b) monoclastic surface; (c) synclastic surface.}
\end{figure}

From  \eqref{nub} we deduce that:
\begin{itemize}
	\item if $0<\nu^{(b)}<1$ the curvatures have opposite signs and hence the plate deforms in an anticlastic surface, Fig. \ref{fig:anticlastic};
	\item if $\nu^{(b)}=0$ the curvature in the $y$ direction is null and hence the plate assumes a cylindrical configuration (monoclastic surface), Fig. \ref{fig:monoclastic};
	\item if $-1<\nu^{(b)}<0$ the curvatures have the same sign and hence the plate deforms in a synclastic surface, Fig. \ref{fig:synclastic}.
\end{itemize}

Equation \eqref{wMx} shows that the extreme values of $\nu^{(b)}$, determined by \eqref{constraint}$_2$, are problematic.
Indeed, one checks that for $\nu^{(b)}=-1$:
$$
Q^{(b)}(\Kb)=\Dc\Big((\tr \Kb)^2-4\det \Kb\Big)=\Dc\Big((K_{11}-K_{22})^2+4K_{12}^2\Big)
$$
and hence
\begin{equation}\label{num-}
\mbox{if $\nu^{(b)}=-1$,}\qquad
Q^{(b)}(\Kb)=0 \iff \Kb=\begin{pmatrix} \alpha & 0 \\ 0 & \alpha \end{pmatrix}
\mbox{ for some $\alpha \in\Ro$}.
\end{equation}
Similarly,
\begin{equation}\label{num+}
\mbox{if $\nu^{(b)}=1$,}\qquad
Q^{(b)}(\Kb)=0 \iff \Kb=\begin{pmatrix} \alpha & \beta \\ \beta & -\alpha \end{pmatrix}
\mbox{ for some $\alpha,\beta\in \Ro$}.
\end{equation}
According to \eqref{num-}
we have that if $\nu^{(b)}=-1$ we may produce synclastic deformations of the plate of type
\begin{equation}\label{energy0-}
w^-(x,y)=\frac{\alpha}2(x^2+y^2),
\end{equation}
with $\alpha\in\Ro$, without paying any energy, {\it i.e.}, $Q^{(b)}(\nabla^2w^-)=0$.

\begin{figure}[h!]
	\centering%
	\subfigure[\label{fig:contourplot}]%
	{\def\svgwidth{.4\textwidth}
		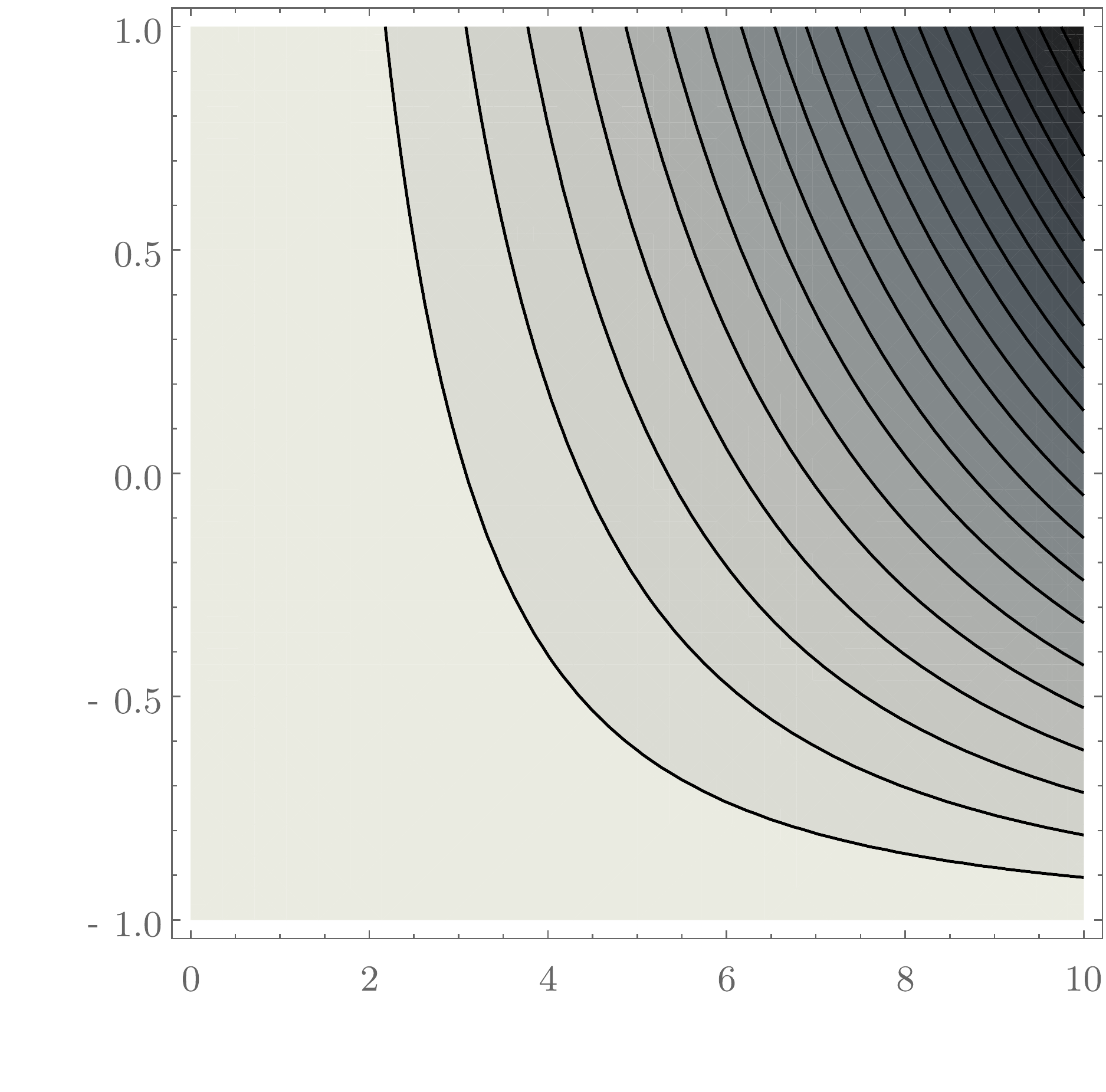}\qquad
	\subfigure[\label{fig:incralpha}]%
	{\includegraphics[scale=0.4]{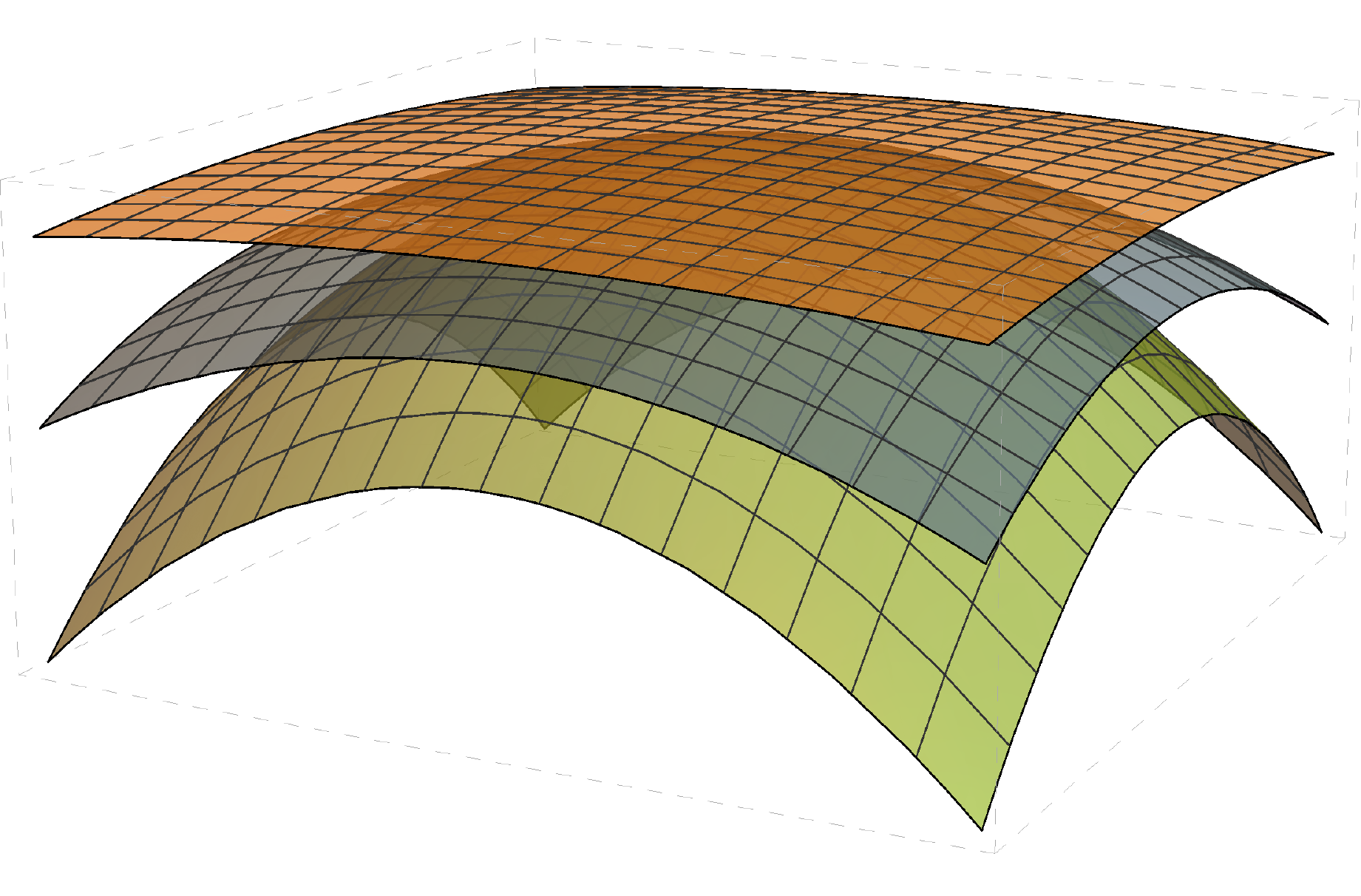}}\qquad
	\caption{(a) Level curves of the bending energy, computed for $w^-$, as a function of the bending Poisson coefficient $\nu^{(b)}$ and $\alpha$; (b) shape of the surface for $\alpha$ increasing.}
\end{figure}

Fig. \ref{fig:contourplot} shows the level curves  of the bending energy $\Uc^{(b)}$, for a fixed $\Dc$, computed for $w^-$ as in \eqref{energy0-}; increasing $\alpha$ corresponds to an higher curvature, see Fig. \ref{fig:incralpha}. When $\nu^{(b)}=-1$,  increasing the curvature does not change the energy.

Similarly, from \eqref{num+}, if $\nu^{(b)}=1$ we may produce anticlastic deformations of the plate of type
\begin{equation}\label{energy0+}
w^+(x,y)=\frac{\alpha}2(x^2-y^2)+\beta xy,
\end{equation}
with $\alpha,\beta\in\Ro$, without paying any energy.

\begin{figure}[h!]
	\centering%
	\subfigure[\label{fig:contourplotp}]%
	{\def\svgwidth{.4\textwidth}
		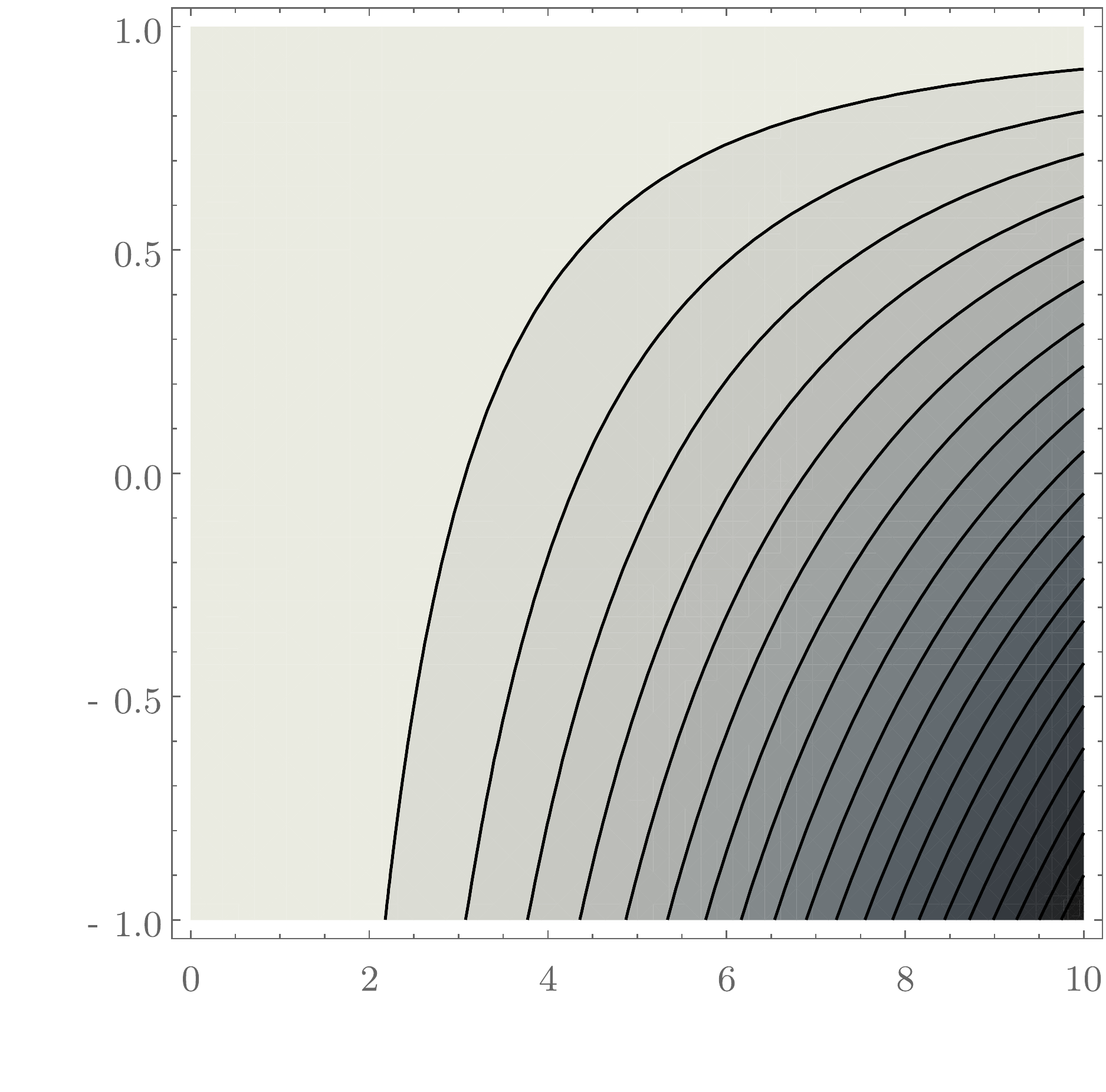}\qquad
	\subfigure[\label{fig:incralphap}]%
	{\includegraphics[scale=0.4]{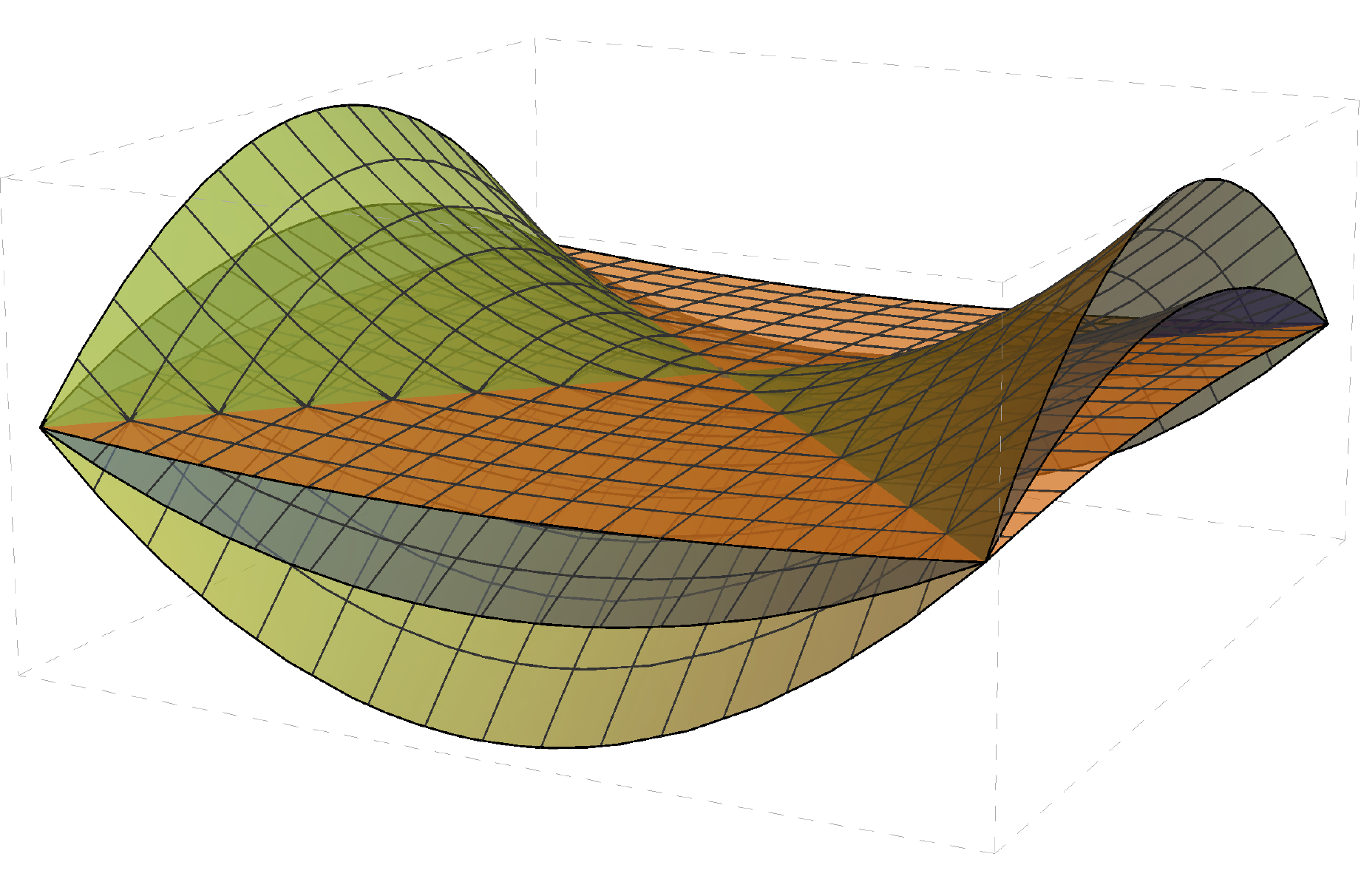}}\qquad
	\caption{(a) Level curves of the bending energy, computed for $w^+$, as a function of  bending Poisson coefficient $\nu^{(b)}$ and $\alpha$; (b) shape of the surface for $\alpha$ increasing and $\beta=0$.}
\end{figure}

Fig. \ref{fig:contourplotp} shows the level curves  of the bending energy $\Uc^{(b)}$, for a fixed $\Dc$, computed for $w^+$ as in \eqref{energy0+}, where we have set $\beta=0$; increasing $\alpha$ corresponds to an higher curvature, see Fig. \eqref{fig:incralphap}. When  $\nu^{(b)}=1$, increasing the curvature does not change the energy.

\subsection{Three-dimensional theories}

We here consider a three-dimensional body with a linear
and isotropic constitutive equation. We denote by
$E$ the Young modulus and by $\nu$ the Poisson coefficient.
It is well known, that these parameters have to satisfy the following
constraints:
\begin{equation}\label{constraint3d}
E>0\qquad\mbox{and}\qquad -1<\nu<\frac 12.
\end{equation}

If the region occupied by the body is the cylinder $\Omega \times (-h/2,h/2)$, with the thickness of the cylinder $h$ much smaller than the diameter of $\Omega$, it seems reasonable and convenient to approximate
the three-dimensional elastic problem with two-dimensional problems. This procedure can be achieved in several ways \cite{PodioGuidugli_1989,Ciarlet_1997,Paroni_2006,Paroni_2006a}, and leads to problems \eqref{Um} and \eqref{Ub}, with
\begin{equation}\label{E3d}
\Ec=\frac{E h}{2(1-\nu^2)}, \qquad \nu^{(m)}=\nu,
\end{equation}
and
\begin{equation}\label{D3d}
\Dc=\frac{E h^3}{12(1-\nu^2)}, \qquad \nu^{(b)}=\nu.
\end{equation}
Hence, we see that if the two-dimensional theories are derived from 3D elasticity,
there is no distinction between membranal 
and bending Poisson coefficients. Moreover, from
\eqref{constraint3d}, \eqref{E3d}, and \eqref{D3d}, we obtain
\begin{equation}\label{EDnu3d}
\Ec>0,\quad\Dc>0,\quad\mbox{and}\qquad -1<\nu^{(m)}=\nu^{(b)}<\frac 12.
\end{equation}
In particular, the constraints imposed by the three-dimensional theory on the Poisson coefficients are
more restrictive than those obtained from the two-dimensional theories.

\vspace{5mm}

In what follows we show how to design a two-dimensional material (hence the constraints \eqref{EDnu3d} do not hold) with a microstructure that leads to a membranal  Poisson coefficient $\nu^{(m)}=1$ and a bending Poisson coefficient $\nu^{(b)}\in [-1,1]$. In particular, by  appropriately tuning two material parameters, it is possible to achieve the extreme values $\nu^{(b)}=-1$ and $\nu^{(b)}=1$.


\section{Microstructure of the metamaterial: geometry and energetics}\label{microstructure}

\begin{figure}[!h]
	\centering
	\subfigure[\label{fig:microstructure}]
	{\includegraphics[scale=1.5]{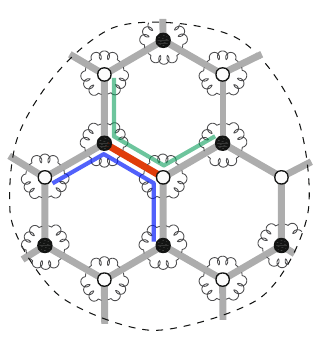}}\qquad\qquad
	\begin{scriptsize}
		\subfigure[\label{fig:lattice}]
		{\def\svgwidth{.4\textwidth}
			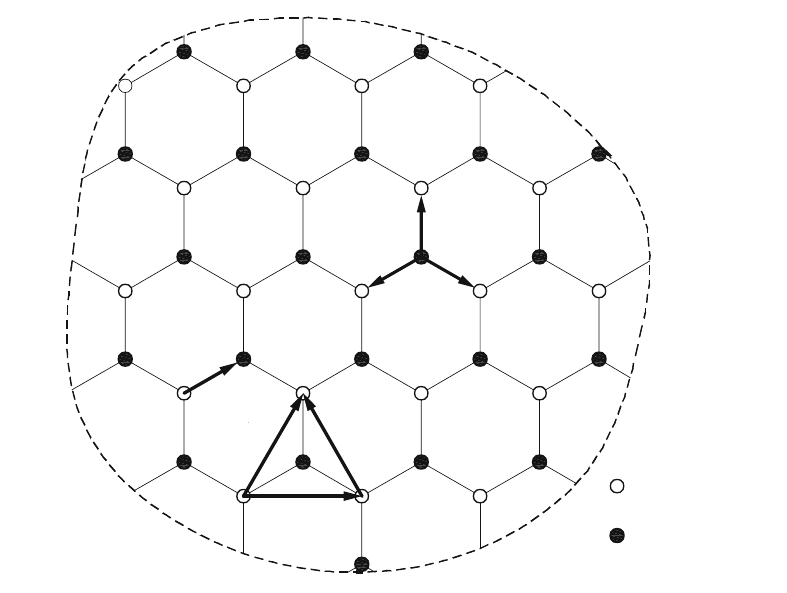}\qquad\qquad
	\end{scriptsize}
	\caption{	(a) Hexagonal lattice with rotational springs. A typical stick (in red) shares two C-shaped sequence of sticks (blue and green). Whenever the dihedral angle spanned by a C-shaped sequence changes, energy is stored.
		(b) Lattice vectors.}
	\label{lattice}
\end{figure}

The lattice material we consider is composed of pin-jointed \textit{rigid sticks} arranged in a hexagonal pattern, pairwise connected by \textit{angular springs}; each stick is the center of two C-shaped sequences of sticks (see Fig. \ref{lattice} a) and we consider  \textit{dihedral springs}, storing energy whenever a change of the dihedral angles  spanned by the C-shaped sequence occurs.

We can describe this array as  a $2$-lattice composed of two Bravais lattices: \begin{equation}\label{eq:KIN_1}
\begin{array}{l} L_1(\ell) = \{ \mathbf{x} \in \mathbb{R}^2:
\mathbf{x} = n^1\ell \db_{1} + n^2\ell \db_{2} \quad \mbox{with} \quad (n^1,
n^2) \in \mathbb{Z}^2 \}, \\ L_2(\ell) = \ell\mathbf{p} + L_1(\ell),
\end{array}
\end{equation}
simply shifted with respect to one another. In Fig. \ref{lattice} the nodes of $L_1$ are represented by blank circles and those of $L_2$ by black spots. In \eqref{eq:KIN_1}, $\ell$ denotes
the lattice size, while
$\ell\db_{\alpha}$ and $\ell\mathbf{p}$ respectively are the {\it lattice
	vectors} and the {\it shift vector}, with:
\begin{equation}\label{eq:KIN_2}
\db_{1} =\sqrt{3}\eb_1, \quad \db_{2} = \frac{\sqrt{3}}{2}\eb_1+
\frac{3}{2}\eb_2, \quad \mbox{and} \quad \mathbf{p} =
\frac{\sqrt{3}}{2}\eb_1+ \frac{1}{2}\eb_2,
\end{equation}
where $\eb_1$ and $\eb_2$ are two orthonormal vectors spanning the plane containing $\Omega$ with unit normal $\eb_3$.
%
%
%
The sides of the hexagonal cells in Fig.~\ref{lattice}  are represented by the
vectors
\begin{equation}\label{eq: KINENER 3}
\pb_\alpha = \db_\alpha - \pb \ \  (\alpha = 1, 2) \quad \mbox{and}
\quad \pb_3 = - \, \pb.
\end{equation}

We consider a bounded lattice of points $\xb^\ell\in L^1(\ell)\cup L^2(\ell)$ contained in a bounded open set $\Omega\subset \mathbb{R}^2$.


We  assume that the stored energy is given by the sum of the following two  terms:

\begin{equation}\label{eq:ENER 1}
\mathcal{U}_\ell^\vartheta = \frac{1}{2} \, \sum_{\mathcal {W}} k^\vartheta (\vartheta-\vartheta^{\rm nat})^2, \quad
\mathcal{U}_\ell^\Theta =  \frac{1}{2} \, \sum_{\mathcal {C}} k^\Theta \, (\delta\Theta)^2\,.
\end{equation}
$\mathcal{U}_\ell^\vartheta $ and $\mathcal{U}_\ell^\Theta$ are the energies of the  the wedges and the dihedra, respectively;  $\vartheta$ denotes the angle between pairs of edges having a lattice point in common and $\Theta$ the  C-dihedral angles between two consecutive wedges. Moreover we consider the possibility that the wedge springs are not stress-free in the reference configuration, i.e. $\vartheta^{\rm nat}=2/3\pi+\delta\vartheta_0$, with $\delta\vartheta_0\neq0$, is the angle at ease between consecutive edges. As a consequence, the wedge energy \eqref{eq:ENER 1}$_1$, up to a constant, takes the form
\begin{equation}\label{utth}
\mathcal{U}_\ell^\vartheta = \tau_{0} \sum_{\mathcal {W}}\delta \vartheta +  \frac{1}{2} \, \sum_{\mathcal {W}} k^\vartheta \, (\delta\vartheta)^2,
\end{equation}
with
\begin{equation}\label{deft0}
\tau_0 := -k^\vartheta \, \delta\vartheta_0
\end{equation}
the {\it wedge self-stress}.

\begin{remark}
	This microstructure is reminiscent of, and inspired by, that of graphene, a well-known two-dimensional material with non-auxetic behavior both in plane and in bending. In graphene,  dihedral energetic contributions are crucial, even if in that case interactions are  more complex and the C-shaped dihedral angles, only, are not sufficient to account for the real behavior. A more general continuum model has been deduced in \cite{Davini_2017} and \cite{Davini_2017b} starting from a discrete energy that includes, in particular, the contributions envisaged in  \eqref{eq:ENER 1}. Some cumbersome computations in common with the present work and here omitted  are fully detailed there.

\end{remark}

We  approximate the strain measures to the lowest order that makes the energy quadratic in the displacement field. It is  then possible to show (cf. \cite{Davini_2017}) that the energy splits into two parts: one ---the membranal energy $\Uc^{(m)}$--- depends on the in-plane displacement and the other ---the bending energy $\Uc^{(b)}$--- is a function of  the out-of-plane displacement $w: L_1(\ell) \cup L_2(\ell)\to\mathbb{R}$. In particular, we have that

\begin{equation}\label{enerdis}
\begin{aligned}
& \mathcal{U}_\ell^{(m)}:=\frac{1}{2} \, \sum_{\mathcal {W}} k^\vartheta \, {(\delta\vartheta^{(1)})}^2, \quad \mathcal{U}_\ell^{(b)}:=\Uc_\ell^{(s)} +\Uc_\ell^{\Theta},
\end{aligned}
\end{equation}
where
\begin{equation}\label{selfen}
\Uc_\ell^{(s)} :=\tau_{0} \sum_{\mathcal {W}}\delta \vartheta^{(2)},
\end{equation}
is the self-energy associated to the self-stress $\tau_0$. In \eqref{enerdis} $\delta\vartheta^{(1)}$ and $\delta\vartheta^{(2)}$ are the first-order and the  second-order variation of the wedge angle with respect to the reference angle $\frac{2}{3}\pi$.

\begin{figure}
	\begin{center}
		\def\svgwidth{.4\textwidth}
		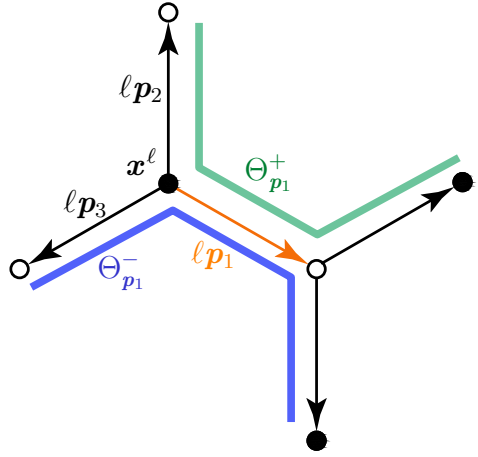
		\caption{Dihedral angles $\Thc_{\pb_1^+}$ (green) and dihedral angles $\Thc_{\pb_1^-}$ (blue) .}
		\label{fig:cell_text}
	\end{center}
\end{figure}

Hereafter we write explicitly the dependence  on $w$ of the strain measures. In particular, by $\Thc_{\pb_i^+}[w](\xb^\ell)$, with $\xb^\ell\in L_2(\ell)$, we denote  the angle corresponding to the dihedron with middle edge $\ell \pb_i$ and oriented as $\eb_3\times \pb_i$,  while $\Thc_{\pb_i^-}[w](\xb^\ell)$ is the angle corresponding to the C-dihedron oriented opposite to $\eb_3\times \pb_i$ (see Fig.~\ref{fig:cell_text} for $i=1$).

In \cite{Davini_2017} it has been shown that the dihedral energy can be written as:
\begin{equation}\label{enC}
\begin{aligned}
\mathcal{U}^{\Theta}_\ell(w):=\frac{1}{2} k^{\Theta} \, \sum_{\xb^\ell\in L_2(\ell)}
\sum_{i=1}^3 \Big(\Thc_{\pb_i^+}[w](\xb^\ell)\Big)^2+ \Big(\Thc_{\pb_i^-}[w](\xb^\ell)\Big)^2,
\end{aligned}
\end{equation}
with the change of dihedral angles given by:
\begin{equation}\label{C-wedge}
\begin{aligned}
\Thc_{\pb_i^+}[w](\xb^\ell)&=+\frac{2\sqrt{3}}{3\ell}[2w(\xb^\ell)-w(\xb^\ell+\ell \pb_{i+1})+w(\xb^\ell+\ell\pb_i-\ell \pb_{i+2})-2w(\xb^\ell+\ell \pb_{i})],\\
\Thc_{\pb_i^-}[w](\xb^\ell)&=-\frac{2\sqrt{3}}{3\ell}[2w(\xb^\ell)-w(\xb^\ell+\ell \pb_{i+2})+w(\xb^\ell+\ell\pb_i-\ell \pb_{i+1})-2w(\xb^\ell+\ell\pb_i)].
\end{aligned}
\end{equation}
To make notation simpler, in the preceding equations we have omitted the symbol $\delta$  in denoting the variations of the dihedral angles; we will do the same  throughout the paper, without any further mention.

Concerning the self-energy,  \eqref{selfen} becomes:
\begin{equation}\label{ens}
\mathcal{U}^{s}_\ell(w):=-\frac{1}{2} \tau_0\,  \Big[\sum_{\xb^\ell\in L_1(\ell)}
\, \Big(\Ths_1[w](\xb^\ell)\Big)^2+\sum_{\xb^\ell\in L_2(\ell)}
\,\Big( \Ths_2[w](\xb^\ell)\Big)^2\Big],
\end{equation}
where:
\begin{equation}\label{s_angles}
\begin{aligned}
\Ths_1[w](\xb^\ell)&=\frac{\sqrt{3\sqrt{3}}}{\ell}\Big(\frac 13 \sum_{i=1}^3w(\xb^\ell-\ell\pb_i)-w(\xb^\ell)\Big), \quad \xb^\ell\in L_1(\ell),\\
\Ths_2[w](\xb^\ell)&=\frac{\sqrt{3\sqrt{3}}}{\ell}\Big(\frac 13 \sum_{i=1}^3w(\xb^\ell+\ell\pb_i)-w(\xb^\ell)\Big), \quad \xb^\ell\in L_2(\ell).
\end{aligned}
\end{equation}
For detailed computations, the reader is referred to \cite{Davini_2017}.

\begin{remark}\label{remark2}
	For the synclastic deformation $w^-$ defined in \eqref{energy0-}, simple computations show that
	the dihedral energy is null:
	\begin{equation}\label{rem2}
	\mathcal{U}^{\Theta}_\ell(w^-)=0
	\end{equation}
	for every $\ell$.
We recall that  in a plate with bending Poisson coefficient $\nu^{(b)}=-1$, the displacement $w^-$ may be produced without paying energy. The same then happens in the discrete case when the self-stress is absent. We will see that this similarity is a sign that the discrete dihedral energy alone will lead, as $\ell$ goes to zero, to a plate model with bending Poisson coefficient equal to $-1$.

Equation \eqref{rem2} holds true because $\Theta[w^-]_{\pb_i^\pm}(\xb^\ell)=0$ for every $i=1,2,3$. As an example, let us consider the case $i=1$.  Recalling that $w^-(\xb)=\frac{\alpha}2(x_1^2+x_2^2)$, we calculate that
		\begin{equation*}
		\begin{aligned}
		&w^-(\xb^\ell)=\frac{\alpha}2({x^\ell_1}^2+{x^\ell_2}^2), \quad w^-(\xb^\ell+\ell\pb_2)=\frac{\alpha}2({x^\ell_1}^2+(x^\ell_2+\ell)^2), \\
		&w^-(\xb^\ell+\ell\pb_1-\ell\pb_3)=\frac{\alpha}2((x^\ell_1+\sqrt{3}\ell)^2+{x^\ell_2}^2),  \\
		&w^-(\xb^\ell+\ell\pb_1)=\frac{\alpha}2((x^\ell_1+\sqrt{3}/2 \ell)^2+(x^\ell_2-\ell/2)^2), \\
		&w^-(\xb^\ell+\ell\pb_3)=\frac{\alpha}2((x^\ell_1-\sqrt{3}/2 \ell)^2+(x^\ell_2-\ell/2)^2), \\
		&w^-(\xb^\ell+\ell\pb_1-\ell\pb_2)=\frac{\alpha}2((x^\ell_1+\sqrt{3}/2 \ell)^2+(x^\ell_2-3/2\ell)^2).
		\end{aligned}
		\end{equation*}
		Hence, by using \eqref{C-wedge}, we find
		$
		\Thc_{\pb_1^+}[w^-](\xb^\ell)=\Thc_{\pb_1^-}[w^-](\xb^\ell)=0.
		$
\end{remark}

\begin{remark}
	        As in Remark \ref{remark2},  for the anticlastic
		deformation $w^+$ defined in \eqref{energy0+} we  have that 
		the self-energy is null:
		\begin{equation*}
		\mathcal{U}^{s}_\ell(w^+)=0
		\end{equation*}
		for every $\ell$. 	
		In the next section we show that the self-energy alone will lead to a plate model with bending Poisson coefficient equal to $1$.
\end{remark}

\section{Poisson coefficients of the metamaterial}\label{Poissons}
\subsection{Membranal Poisson coefficient}
As stated in \eqref{enerdis}$_1$, the membranal energy depends just on the first-order variation of the wedge angle, while the dihedral energy and the self-energy do not play any role. The microstructure is then  equivalent to that of rigid hexagonal pin-jointed sticks with rotational springs. It is known that, in this case, the membranal  Poisson coefficient is
\begin{equation}
\nu^{(m)}=1,
\end{equation}
see e.g. \cite{Gibson_1982,Gibson_1999, Berinskii_2016}.

\begin{remark}
Indeed, according to an asymptotic model for the in-plane deformations of a monolayer graphene sheet in which the axial deformability of the sticks is taken into account, see \cite{Davini_2014, Berinskii_2016}, the membranal Poisson coefficient is given by the formula
		\begin{equation}\label{membr}
		\nu^{(m)}=\frac{k^l-6 k^\vartheta}{k^l+18 k^\vartheta},
		\end{equation}
		where $k^l$ is the axial stiffness.  On tuning the value of the stiffnesses, it is then possible to get an in-plane non auxetic material (for large $k^\ell$) or  an in-plane auxetic material (for large $k^\vartheta$); for instance, if $k^\vartheta\to\infty$, the membranal Poisson coefficient $\nu^{(m)}\to-1/3$ and the material is auxetic.
\end{remark}

\subsection{Bending Poisson coefficient}
The continuous bending energy can be deduced from the discrete energy $\Uc_\ell^{(b)}$ by  letting the lattice size $\ell$ go to zero so that
$L_1(\ell)\cup L_2(\ell)$ invades $\Omega$.  The limit  turns out to be
\begin{align}\label{entot}
\mathcal{U}^{(b)}_0(w):=&\Uc_0^\Theta+\Uc_0^{(s)}\\
=&\frac 12 \Bigg(\frac{2\sqrt{3}}{3} k^{\Theta}-\frac{\tau_0}2\Bigg) \int_\Omega(\Delta w)^2-2\left( 2+\frac{6 \tau_0}{4\sqrt{3}k^\Theta-3\tau_0} \right)\det\nabla^2w\,d\xb.
\end{align}
For detailed computations, the reader is referred to \cite{Davini_2017,Davini_2017b}, where,  in a different context, a more general case is treated.
By comparing \eqref{entot} and \eqref{Ub}, we find that the bending Poisson coefficient is 
\begin{equation}\label{nubb}
\nu^{(b)}=-\left(1+\frac{6 \tau_0}{4\sqrt{3}k^\Theta-3\tau_0}\right)=-\frac{4\sqrt{3}k^\Theta+3\tau_0}{4\sqrt{3}k^\Theta-3\tau_0},
\end{equation}
and the bending stiffness
\begin{equation}\label{bendstiff}
\Dc=\frac{2\sqrt{3}}{3}k^\Theta-\frac{\tau_0}{2}.
\end{equation}

The expressions \eqref{nubb} and \eqref{bendstiff}  explicitly show the   origin of the continuum material parameters ruling the bending behavior of the metamaterial we consider. In particular, the selfstress $\tau_0$ and the dihedral stiffness $k^\Theta$ completely determine the response to bending. On tuning their magnitude, it is possible to design a metamaterial that has non-auxetic in-plane behavior, but it is able to attain all the admissible values of $\nu^{(b)}$.
It is worth noticing that none of the parameters that determines the membranal Poisson coefficient \eqref{membr}  affects the bending Poisson coefficient, revealing that these two parameters are indeed independent. As a consequence,  a wide variety of mechanical metamaterials can be designed with in-plane auxetic and out-of-plane non-auxetic properties, or vice-versa.

With reference to the out-of-plane behavior, the following cases can be obtained:
	\begin{itemize}
		\item[(i)]  $\nu^{(b)}=-1, \quad\mbox{if the self-stress is null}.$
		
		\noindent The material has then an auxetic out-of-plane behavior, while keeping a non-auxetic in-plane behavior.
		\item[(ii)] 
		$
		\nu^{(b)}=1,\quad\mbox{if the dihedral stiffness is null}.
		$
		
		\noindent The material is then non-auxetic both in plane and in bending.
		
		\item[(iii)] Let $\tau_0\neq 0$, $k^\Theta\neq 0$, and set $\tau_0=-\xi k^\Theta$, $\xi>0$. Then, the bending Poisson coefficient can be written in the form
		
		$$
		\nu^{(b)}=-\frac{4\sqrt{3}-3\xi}{4\sqrt{3}+3\xi},
		$$
		and we have
		$$
		\nu^{(b)}=0,\quad\mbox{if $\xi:=-\displaystyle\frac{\tau_0}{k^\Theta}=4\sqrt{3}/3$}.
		$$
	\end{itemize}

\begin{figure} [h!]
	\begin{center}
		\begin{scriptsize}
			\def\svgwidth{.5\textwidth}
			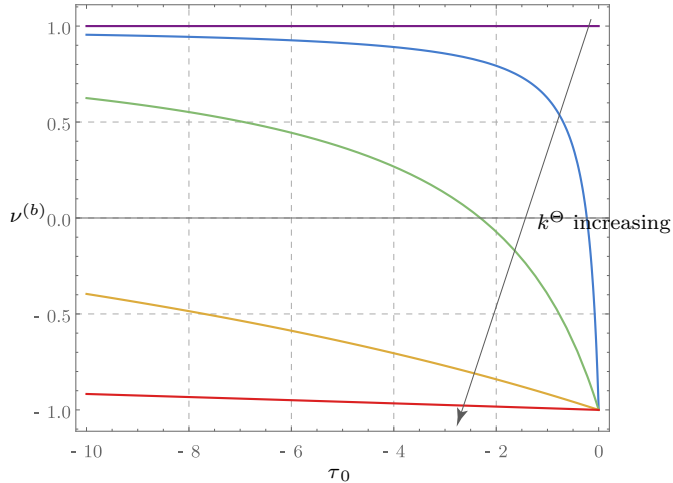
		\end{scriptsize}
		\caption{The bending Poisson coefficient in terms of $\tau_0$, for different values of $k^\Theta$.}
		\label{fig:nutk}
	\end{center}
\end{figure}

In Fig. \ref{fig:nutk} the bending Poisson coefficient is plotted in terms of $\tau_0$, for fixed values of $k^\Theta$. For $k^\Theta>0$, the self-stress yields an increment  of $\nu^{(b)}$ and then the material tends more and more to form an  anticlastic  surface.

\begin{figure}[h!]
	\begin{center}
		\begin{scriptsize}
			\def\svgwidth{.5\textwidth}
			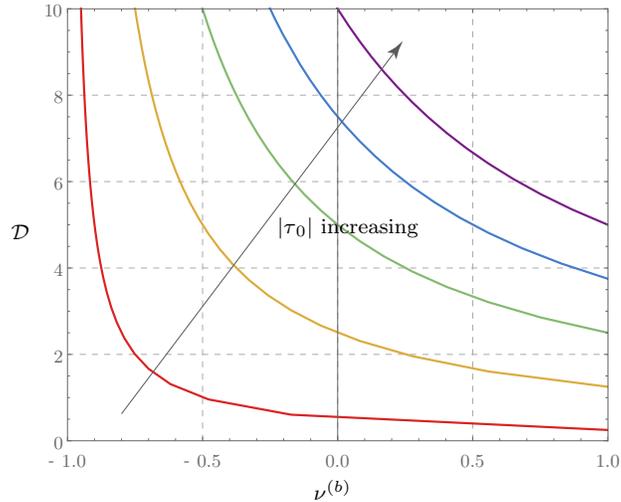
				\end{scriptsize}
		\caption{The bending stiffness $\Dc$ as a function of $\nu^{(b)}$, for different values of $\tau_0$.}
		\label{fig:Dnu} 		
	\end{center}
\end{figure}

Unlike the membranal Poisson coefficient, tuning $\nu^{(b)}$ has relevant consequences on the  bending stiffness $\Dc$, as revealed by Fig. \ref{fig:Dnu}, which illustrates the dependence of $\Dc$ on $\nu^{(b)}$, for different values of the selfstress $\tau_0$. In particular, we find that increasing $\nu^{(b)}$ produces a more and more accentuated monotonic softening effect in the bending stiffness, which approaches the value $-\tau_0/2$. On the contrary, increasing  the self-stress mitigates the softening.

%
%
%

\section{Numerical results}\label{numerical}
In this section we collect some numerical results validating our theory. 
We  solved a set of discrete problems with different  values of $\tau_0$ and $k^\Theta$ and lattice length $\ell$, with the purpose to compare the  continuum material parameter $\nu^{(b)}$ with the predictions computed in the discrete framework.

In \cite{Favata_2014}  a theory of self-stressed elastic molecular structures has been presented, and an \textit{ad hoc} computer code has been developed. We here have specialized that code to the specific geometry considered and generalized the strain measures adopted, in order to include  the dihedral contribution, formerly disregarded.

\begin{figure}[h!]
	\centering
	\def\svgwidth{.5\textwidth}
	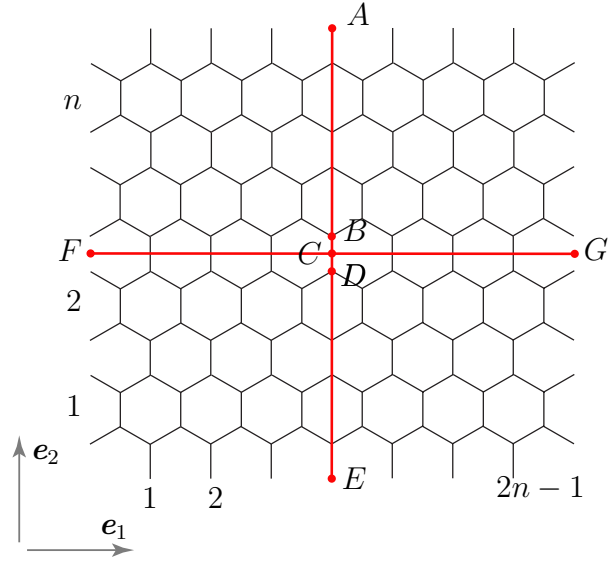
	\caption{Top view of the lattice.}
	\label{fig:hexagons}
\end{figure}

We considered a rectangular sheet  composed of $2n-1$ cells in $\eb_1$ direction and $n$ cells in $\eb_2$ direction (see Fig. \ref{fig:hexagons}). In order to simulate bending, on two opposite sides of the sheet we applied a system of forces statically equivalent to distributed couples. In the simulations, the dimension of the sheet has been kept constant, while  $n$ has been progressively increased.

The comparison between discrete and continuum parameters has been achieved as follows. For a fixed $n$, the nodal displacements  $w$ of the points $A$, $B$, $D$ and $E$ have been determined. The polygonal chain has been  interpolated  by a second-order polynomial function $w_y(y)$, which approximates $w(a/2,y)$. Similarly it has been done with the displacements of the points $F$, $G$, $C$, interpolated by a second-order polynomial function  $w_x(x)$, which approximates $ w(x,b/2)$.
On recalling \eqref{nub}, we have defined the discrete bending Poisson coefficient as
\begin{equation}\label{nub1}
\nu^{(b)}_d:=-\frac{\partial_{yy}w_y(y)}{\partial_{xx}w_x(x)}.
\end{equation}
This value is then compared with the result inferred from \eqref{nubb}.  In table \ref{tab:nub} we have collected the numerical results obtained. We notice that, increasing $n$, the value provided by \eqref{nub1}   better and better approximates the continuum limit \eqref{nubb}.

\begin{center}
	\begin{table}[ht]
		\begin{center}\label{tab:nub}
			\renewcommand{\arraystretch}{2}
			\begin{tabular}{ccc}
				\toprule
				$n=6$ & $n=12$ & $n=32$\\
				\hline
				\multicolumn{3}{c}{$\nu^{(b)}=-0.999360$}\\	
				$-0.999191$&  $-0.999299$ &$-0.999332$ \\
				\hline
				\multicolumn{3}{c}{$\nu^{(b)}=-0.5$}\\	
				$-0.450934$&  $-0.475012$ &$-0.490444$ \\
				\hline
				\multicolumn{3}{c}{$\nu^{(b)}=0$}\\	
				0.035461& 0.017124  &0.006322 \\
				\hline
					\multicolumn{3}{c}{$\nu^{(b)}=0.5$}\\	
					0.513098& 0.505313  & 0.50169 \\
					\hline
				\multicolumn{3}{c}{$\nu^{(b)}=0.999600$}\\	
				0.999577 & 0.999584  & 0.999598\\
				\bottomrule
			\end{tabular}
		\end{center}
		\caption{Comparison between the bending Poisson coefficient $\nu^{(b)}$ computed as in the continuum limit \eqref{nubb} and the  discrete evaluation resulting from \eqref{nub1}.}
	\end{table}
\end{center}

Numerical results show that the evaluation of the overall response to bending is very well approximated by the formula \eqref{nubb}, even in the case of a small number of cells. This leads to conclude that, in the design process of such a metamaterial, the lattice size $\ell$ is not crucial. It is worth noticing that  the percentage error  between discrete and continuum evaluation is less relevant when the extreme cases $\nu^{(b)}=\pm 1$ are considered.

\begin{figure}[H]
	\begin{scriptsize}
		\begin{center}
			\subfigure[\label{fig:n6xz}]%
			{\def\svgwidth{.4\textwidth}
				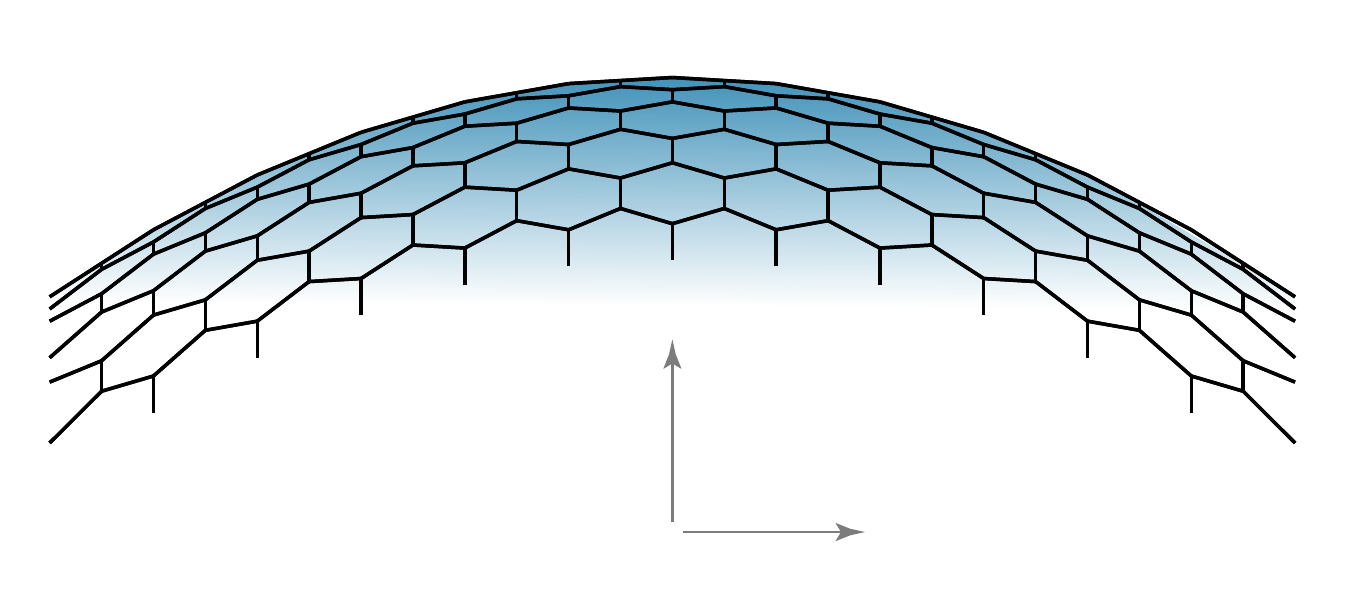}\qquad\qquad
			\subfigure[\label{fig:n6yz}]%
			{\def\svgwidth{.4\textwidth}
				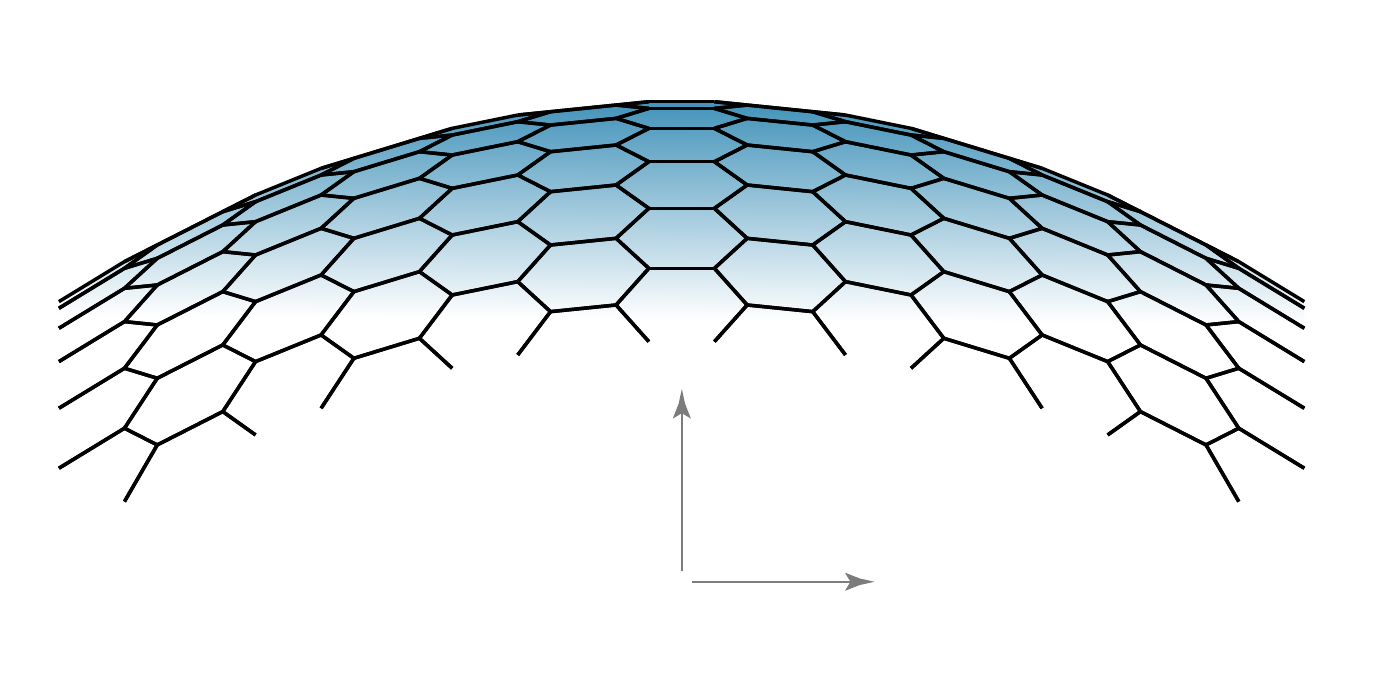}\qquad\qquad
			\subfigure[\label{fig:n6ass}]%
			{\includegraphics[width=0.5\linewidth]{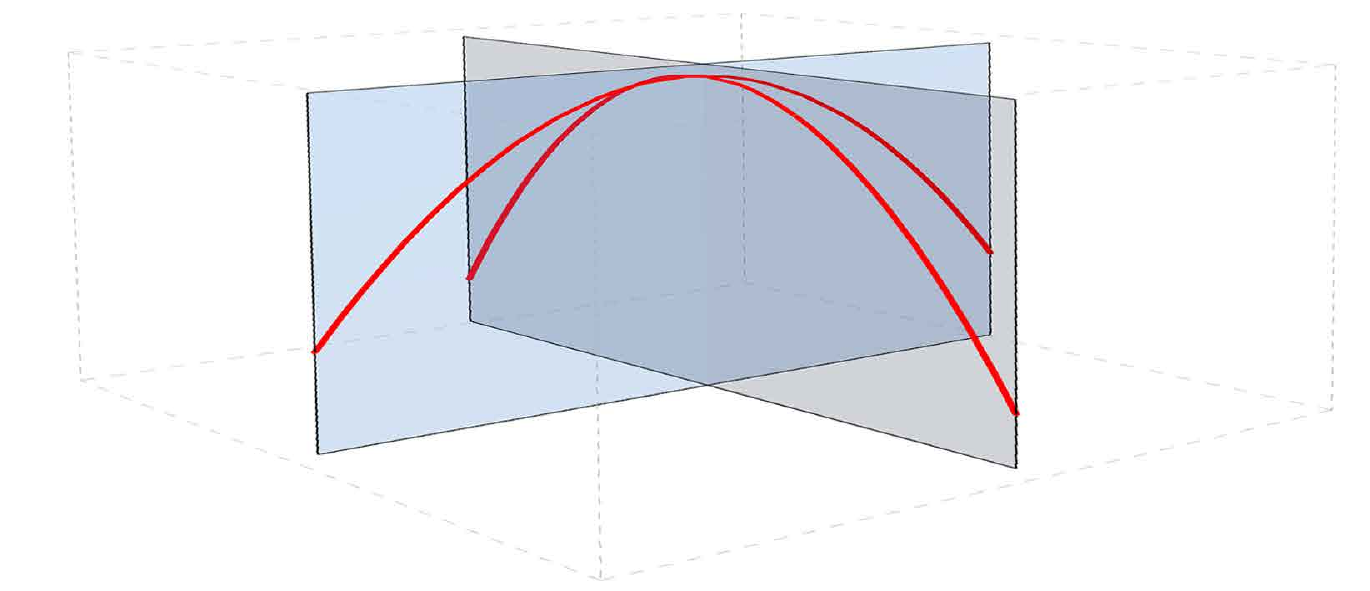}}\qquad\qquad
		\end{center}
	\end{scriptsize}	
	\caption{Case $\nu^{(b)}\simeq-1$. (a) View of the deformed surface in the plane $x-z$; (b) View of the deformed surface in the plane $y-z$; (c) perspective view of the mid-lines. }
\end{figure}

\begin{figure}[H]
	\begin{scriptsize}
		\begin{center}
			\subfigure[\label{fig:n6xz0}]%
			{\def\svgwidth{.4\textwidth}
				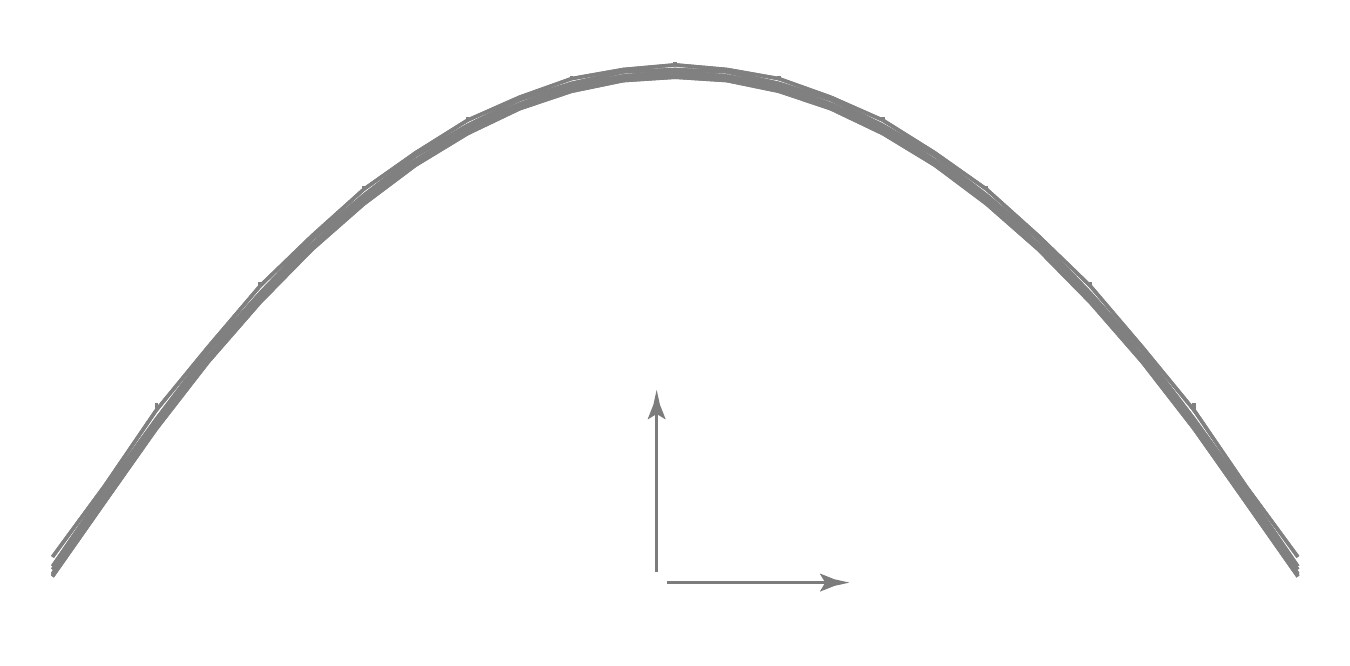}\qquad\qquad
			\subfigure[\label{fig:n6yz0}]%
			{\def\svgwidth{.4\textwidth}
				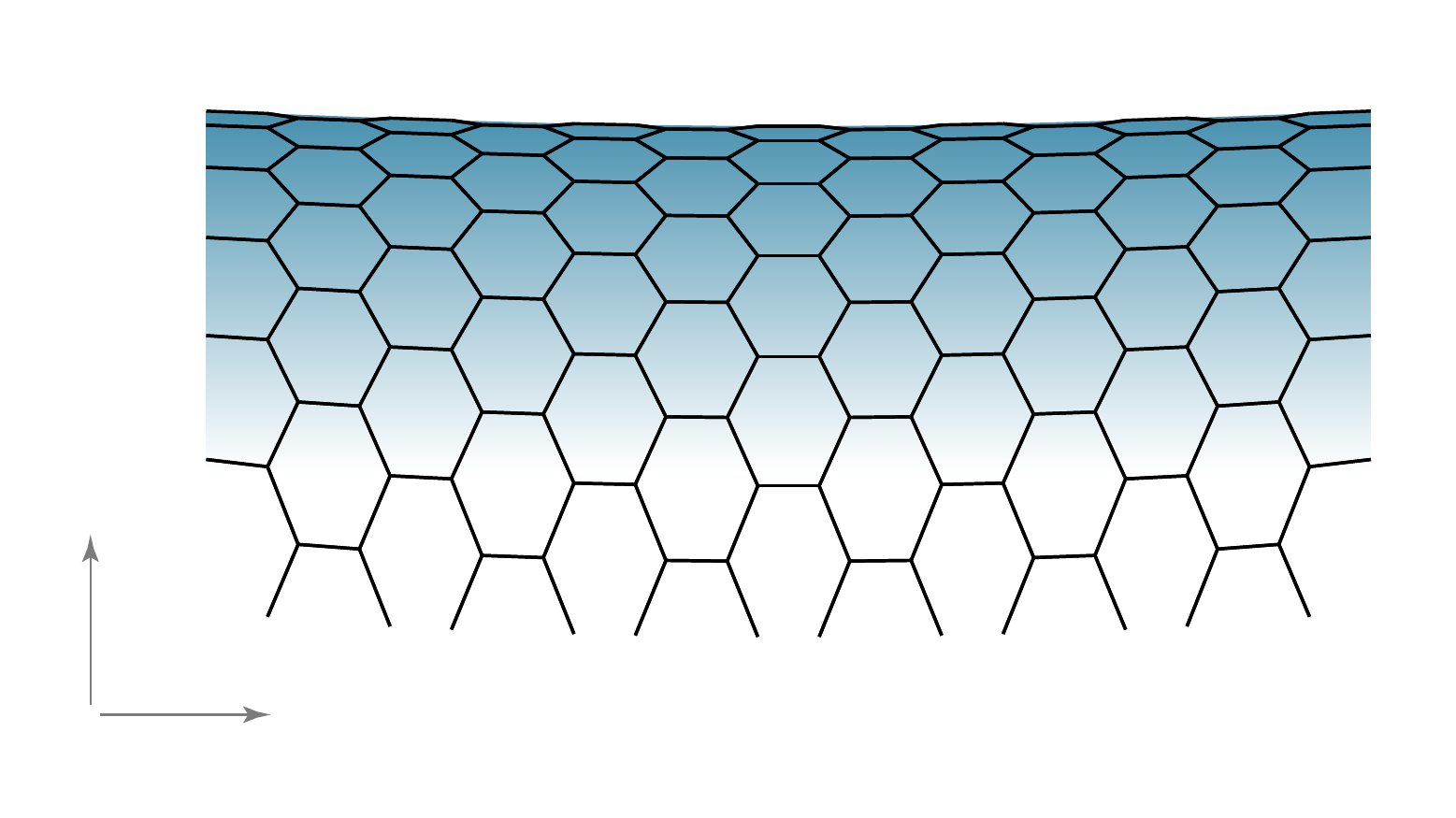}\qquad\qquad
			\subfigure[\label{fig:n6ass0}]%
			{\includegraphics[width=0.4\linewidth]{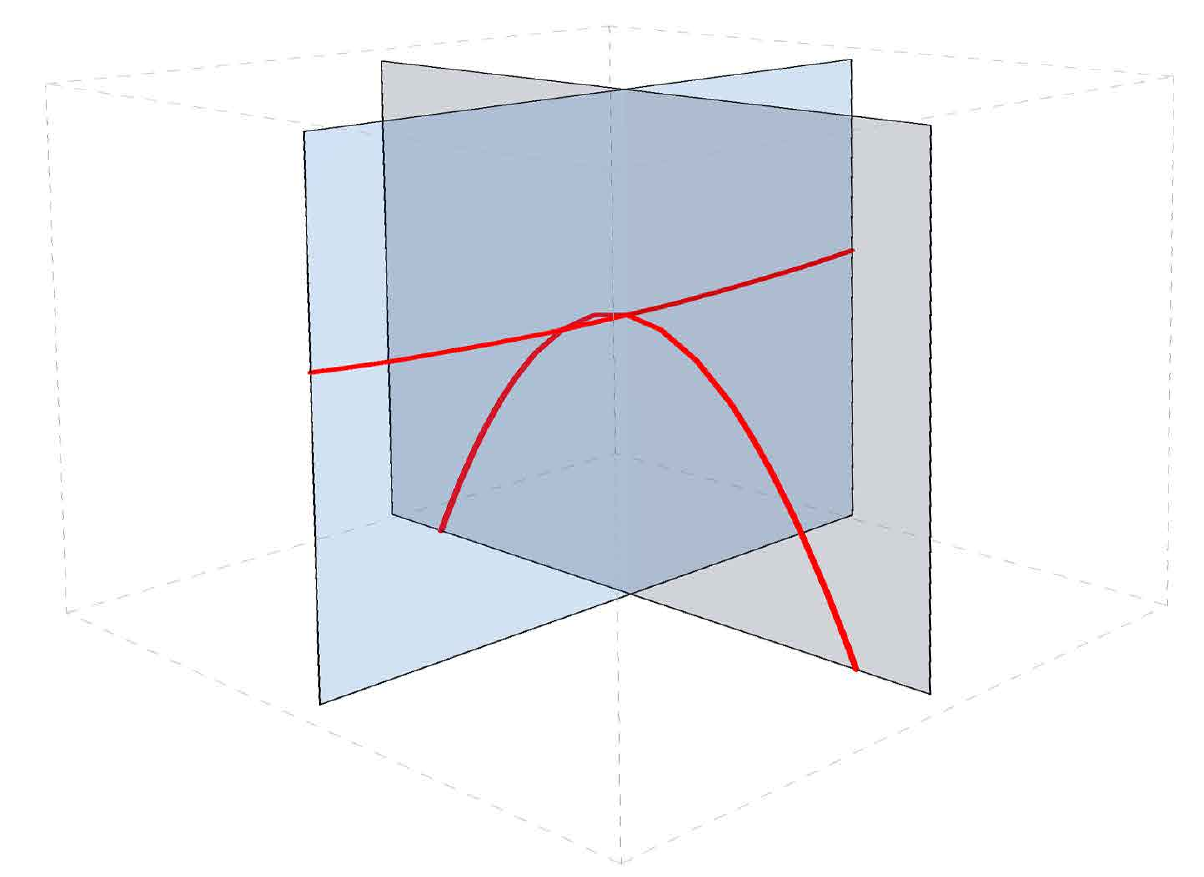}}\qquad\qquad
		\end{center}
	\end{scriptsize}	
	\caption{Case $\nu^{(b)}\simeq0$. (a) View of the deformed surface in the plane $x-z$; (b) View of the deformed surface in the plane $y-z$; (c) perspective view of the mid-lines. }
	\label{fig:n60}
\end{figure}

\begin{figure}[H]
	\begin{scriptsize}
		\begin{center}
			\subfigure[\label{fig:n6xz1}]%
			{\def\svgwidth{.4\textwidth}
				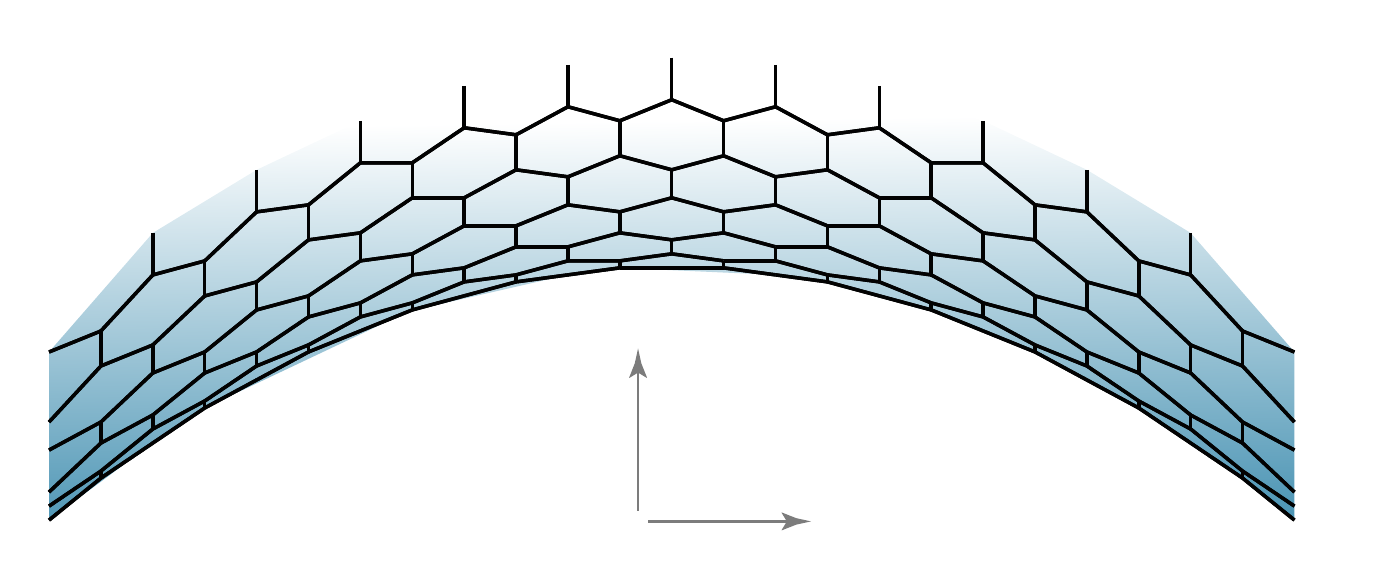}\qquad\qquad
			\subfigure[\label{fig:n6yz1}]%
			{\def\svgwidth{.4\textwidth}
				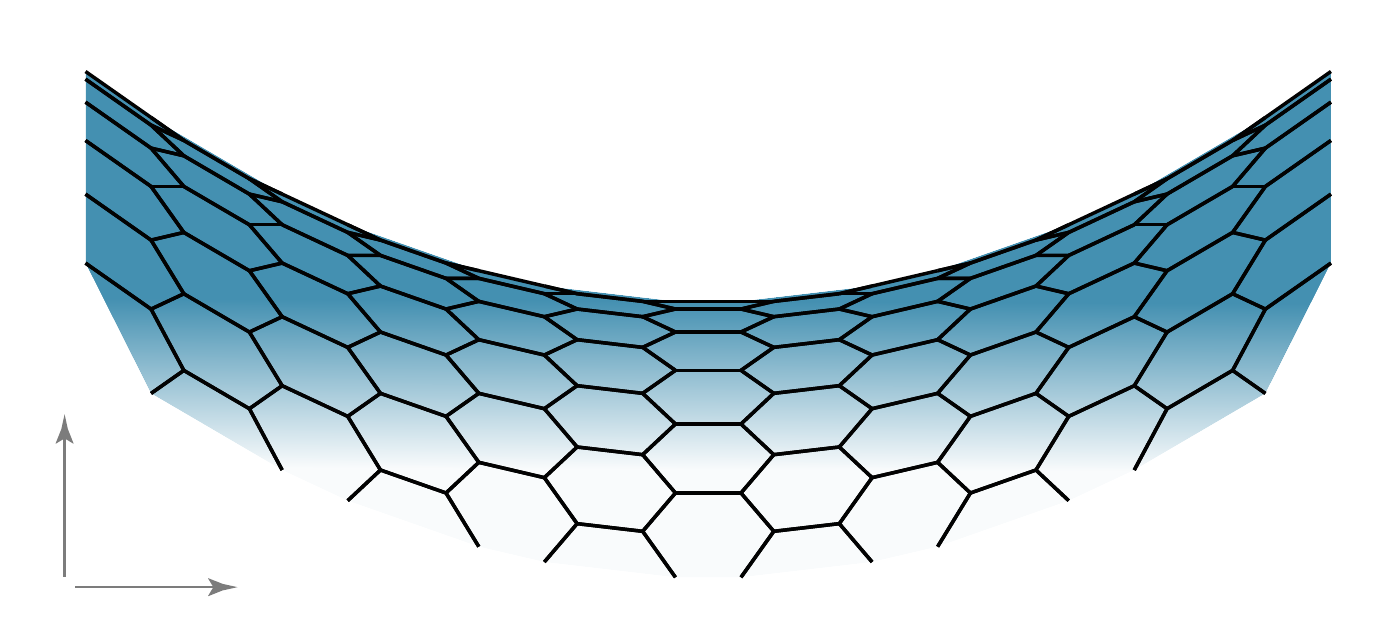}\qquad\qquad
			\subfigure[\label{fig:n6ass1}]%
			{\includegraphics[width=0.4\linewidth]{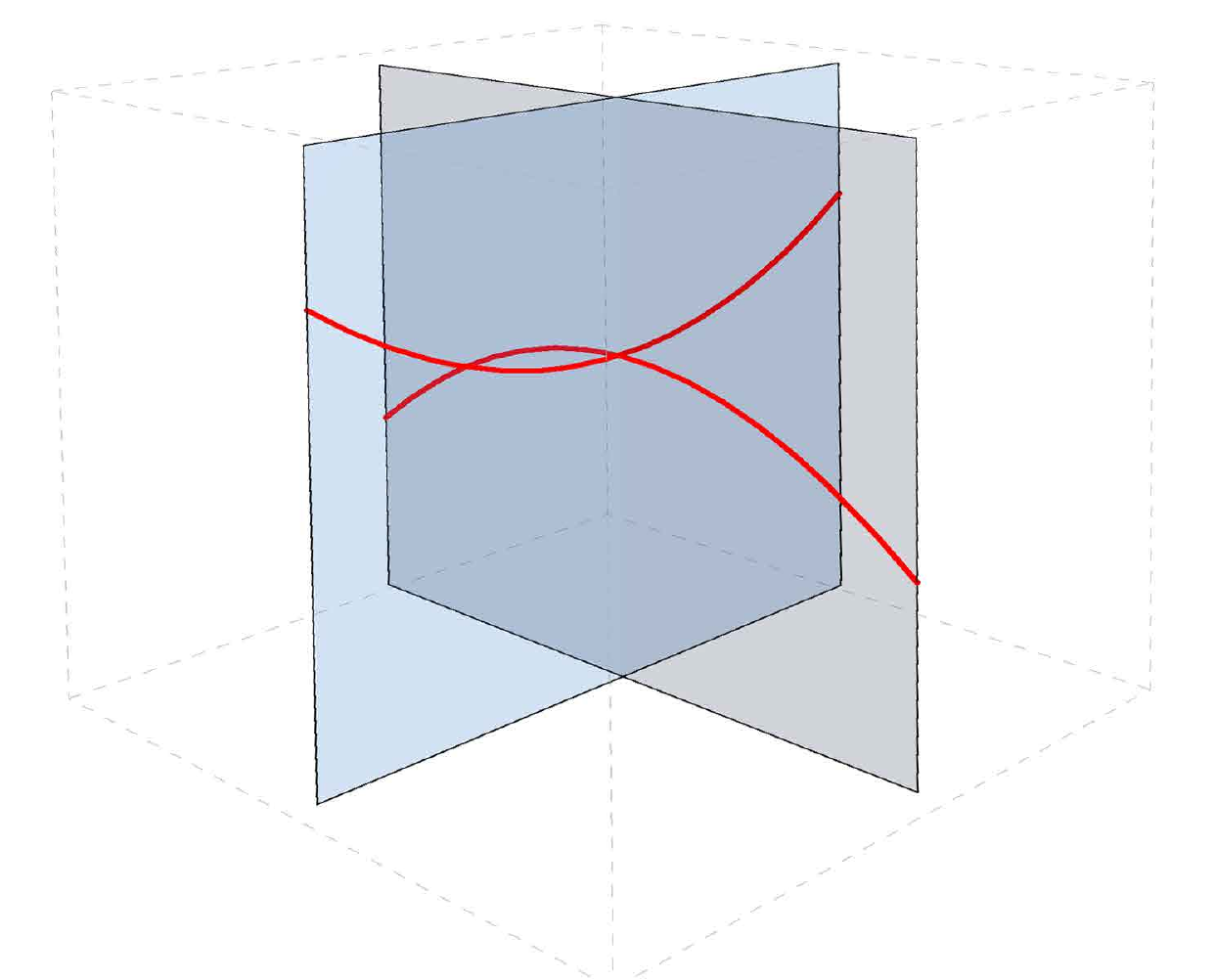}}\qquad\qquad
		\end{center}
	\end{scriptsize}	
	\caption{Case $\nu^{(b)}\simeq1$. (a) View of the deformed surface in the plane $x-z$; (b) View of the deformed surface in the plane $y-z$; (c) perspective view of the mid-lines. }
	\label{fig:n61}
\end{figure}

For $n=6$, we have represented the deformed shape of the sheet. 
Figs. \ref{fig:n6xz} and \ref{fig:n6yz} show two views in the plane $x-z$ and $y-z$, respectively, when $\nu^{(b)}\simeq-1$; the surface is clearly synclastic. The red lines in Fig.  \ref{fig:hexagons} are finally represented in \ref{fig:n6ass}, in the deformed configuration.
The case $\nu^{(b)}\simeq0$ is represented in Fig. \ref{fig:n60}. The surface is essentially monoclastic.
Finally, the case $\nu^{(b)}\simeq1$ is represented in Fig. \ref{fig:n61}, when the surface is  anticlastic.

\section*{Acknowledgments}	
The authors gratefully acknowledge Michele Brun for partly inspiring, with a nice seminar, this work.
AF acknowledges the financial support of Sapienza University of Rome (Progetto d'Ateneo 2016 --- ``Multiscale Mechanics of 2D Materials: Modeling and Applications''). 

\bibliographystyle{plain}
\bibliography{bibtex}

\addcontentsline{toc}{section}{References}

\end{document}